%% file: main.tex
\documentclass[sigconf,screen,nonacm]{acmart}

\usepackage[inline]{enumitem}
\usepackage{acronym}
\usepackage{multirow}
\usepackage{pifont}% http://ctan.org/pkg/pifont
\usepackage[skip=0pt]{caption}

\acrodef{DR}{dense retrieval}
\acrodef{LM}{language model}
\acrodef{MIPS}{maximal inner product search}

\AtBeginDocument{%
  \providecommand\BibTeX{{%
    \normalfont B\kern-0.5em{\scshape i\kern-0.25em b}\kern-0.8em\TeX}}}

\newcommand{\code}[1]{{\ttfamily#1}}

\newcommand{\header}[1]{\vspace*{1mm}\noindent\textbf{#1.}}

\definecolor{lightgoldenrodyellow}{rgb}{0.98, 0.98, 0.82}
\newcommand{\inyellow}[1]{{\setlength\fboxsep{0pt}\colorbox{lightgoldenrodyellow}{#1}}}

\definecolor{lightblue}{rgb}{0.68, 0.85, 0.9}

\definecolor{vlgray}{rgb}{0.87,0.87,0.87}
\newcommand{\ingray}[1]{{\setlength\fboxsep{0pt}\colorbox{vlgray}{#1}}}

\newcommand{\cmark}{\ding{51}}%
\newcommand{\xmark}{\ding{55}}%

\DeclareMathOperator*{\argmax}{arg\,max}

\begin{document}

\title{Learning to Tokenize for Generative Retrieval}

\author{\textbf{Weiwei Sun}\textsuperscript{\rm 1}, \textbf{Lingyong Yan}\textsuperscript{\rm 2}, \textbf{Zheng Chen}\textsuperscript{\rm 1}, \textbf{Shuaiqiang Wang}\textsuperscript{\rm 2}, \textbf{Haichao Zhu}\textsuperscript{\rm 2}\\\textbf{Pengjie Ren}\textsuperscript{\rm 1}, \textbf{Zhumin Chen}\textsuperscript{\rm 1}, \textbf{Dawei Yin}\textsuperscript{\rm 2}, \textbf{Maarten de Rijke}\textsuperscript{\rm 3}, \textbf{Zhaochun Ren}\textsuperscript{\rm 1}\\
\textsuperscript{\rm 1}Shandong University \quad \textsuperscript{\rm 2}Baidu Inc \quad \textsuperscript{\rm 3}University of Amsterdam\\
\small
\{sunnweiwei,lingyongy,shqiang.wang\}@gmail.com,\{202000130223,renpengjie,chenzhumin,zhaochun.ren\}@sdu.edu.cn
hczhu@ir.hit.edu.cn,yindawei@acm.org,m.derijke@uva.nl\\
}

\renewcommand{\shortauthors}{Sun.}

\begin{abstract}
Conventional document retrieval techniques are mainly based on the index-retrieve paradigm. 
It is challenging to optimize pipelines based on this paradigm in an end-to-end manner. 
As an alternative, generative retrieval represents documents as identifiers (docid) and retrieves documents by generating docids, enabling end-to-end modeling of document retrieval tasks. 
However, it is an open question how one should define the document identifiers. 
Current approaches to the task of defining document identifiers rely on fixed rule-based docids, such as the title of a document or the result of clustering BERT embeddings, which often fail to capture the complete semantic information of a document. 

We propose \textsc{GenRet}, a document tokenization learning method to address the challenge of defining document identifiers for generative retrieval.
\textsc{GenRet} learns to tokenize documents into short discrete representations (i.e., docids) via a discrete auto-encoding approach.
Three components are included in \textsc{GenRet}: 
(i) a tokenization model that produces docids for documents; 
(ii) a reconstruction model that learns to reconstruct a document based on a docid; and 
(iii) a sequence-to-sequence retrieval model that generates relevant document identifiers directly for a designated query.
By using an auto-encoding framework, \textsc{GenRet} learns semantic docids in a fully end-to-end manner, where the produced docids can be reconstructed back to the original documents to ensure their semantics.
We also develop a progressive training scheme to capture the autoregressive nature of docids and to stabilize training.

We conduct experiments on the NQ320K, MS MARCO, and BEIR datasets to assess the effectiveness of \textsc{GenRet}.
\textsc{GenRet} establishes the new state-of-the-art on the NQ320K dataset.
Especially, compared to generative retrieval baselines, \textsc{GenRet} can achieve significant improvements on the unseen documents (e.g., at least +14\% relative improvements in terms of R@1).
Furthermore, GenRet can better represent and retrieve documents that have not been seen during the training phase compared to previous rule-based tokenization methods. 
\textsc{GenRet} also outperforms comparable baselines on MS MARCO and BEIR, demonstrating the method's generalizability.\footnote{Preprint. Work in progress.}

\end{abstract}

\maketitle
\acresetall

\input{sections/01-Introduction}
\input{sections/03-Preliminary}
\input{sections/04-Method}
\input{sections/05-Experiment}
\input{sections/06-Results}
\input{sections/02-RelatedWork}

\input{sections/07-Conclusion}

\bibliography{references}
\balance
\bibliographystyle{ACM-Reference-Format}
\input{appendix.tex}

\end{document}

%% file: sections/01-Introduction.tex
\section{Introduction}

Document retrieval plays an essential role in web search applications and various downstream knowledge-intensive tasks, such as question-answering and dialogue systems as it is aimed on identifying relevant documents to satisfy users' queries.
Most traditional document retrieval approaches apply \emph{sparse retrieval} methods, which rely on building an inverted index with term matching metrics such as TF-IDF~\citep{Robertson1997OnRW}, query likelihood~\citep{Lafferty2001DocumentLM}, or BM25~\citep{Robertson2009ThePR}.
The term matching metrics, however, often suffer from a lexical mismatch~\citep{Lin2020PretrainedTF}.

\begin{figure}[h]
 \centering
\includegraphics[width=1\columnwidth]{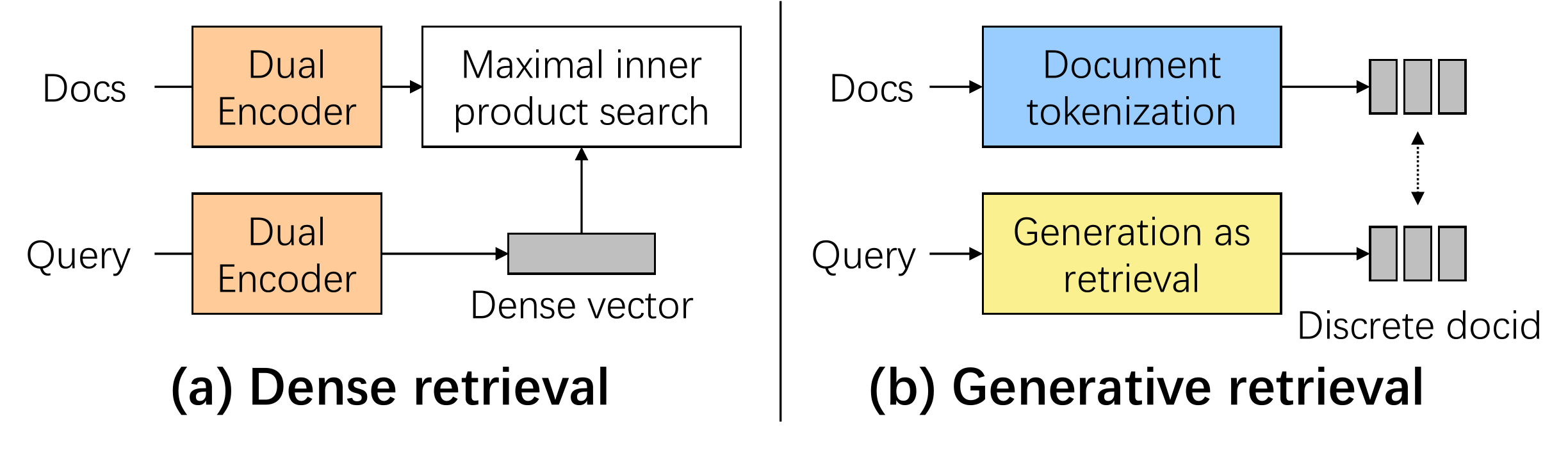} 
%\vspace*{-7mm}
\caption{Two types of document retrieval models: (a) \emph{Dense retrieval} encodes queries and documents to dense vectors and retrieves documents by \ac{MIPS}; (b) \emph{Generative retrieval} tokenizes documents as docids and autoregressively generates docids as retrieval results.
}
\label{fig:background}
\end{figure}

Major progress has recently been made in \acfi{DR} models due to advances in pre-trained \acp{LM}~\citep{Gillick2018EndtoEndRI,Karpukhin2020DensePR,Xiong2020ApproximateNN,Ni2021LargeDE}.
As illustrated in Figure~\ref{fig:background}~(a), \ac{DR} methods learn dense representations of both queries and documents using dual encoders, and subsequently retrieve relevant documents using \ac{MIPS}~\citep{Karpukhin2020DensePR,Johnson2017BillionScaleSS}.
\ac{DR} methods are able to address the lexical mismatch issue with state-of-the-art performance on various retrieval tasks~\citep{Liu2021PretrainedLM,Neelakantan2022TextAC}.

Despite their success, \ac{DR} approaches face two main limitations~\citep{DeCao2020AutoregressiveER,Metzler2021RethinkingSM}:
\begin{enumerate*}[label=(\roman*)]
    \item \ac{DR} models employ an index-retrieval pipeline with a fixed search procedure (\ac{MIPS}), making it difficult to jointly optimize all modules in an end-to-end way; and
    \item The learning strategies (e.g., contrastive learning~\citep{Karpukhin2020DensePR}) are usually not consistent with the pre-training objectives, such as the next token prediction~\citep{Brown2020LanguageMA}, which makes it hard to leverage knowledge in pre-trained LMs~\citep{Bevilacqua2022AutoregressiveSE}.
\end{enumerate*}

\header{Generative retrieval}
Recently, \emph{generative retrieval} has emerged as a new paradigm for document retrieval~\citep{DeCao2020AutoregressiveER,Tay2022TransformerMA,Bevilacqua2022AutoregressiveSE,Wang2022ANC,Zhou2022UltronAU,Zhuang2022BridgingTG}.
As illustrated in Figure~\ref{fig:background}~(b), generative retrieval models directly generate a ranked list of document identifiers (docids) for a given query using generative \acp{LM}.
Specifically, there are two main steps involved in generative retrieval models:
\begin{enumerate*}[label=(\roman*)]
    \item \emph{Document tokenization}, where each document in the corpus is tokenized as a sequence of discrete characters, i.e., docids, and
    \item \emph{Generation as retrieval}, where the docids of relevant documents are output by autoregressively decoding for a given query.
\end{enumerate*}
Unlike \ac{DR}, the generative paradigm presents an end-to-end solution for document retrieval tasks~\citep{Tay2022TransformerMA}. 
It also offers a promising approach to better exploit the capabilities of recent large \acp{LM}~\citep{Bevilacqua2022AutoregressiveSE,Wang2022ANC}.

\header{Learning to tokenize documents}
Document tokenization plays a crucial role in generative retrieval, as it defines how the document is distributed in the semantic space~\citep{Tay2022TransformerMA}.
And it is still an open problem how to define the document identifiers. 
Most previous generative methods tend to employ rule-based document tokenizers, such as generating titles or URLs~\citep{DeCao2020AutoregressiveER,Zhou2022UltronAU}, or clustering results from off-the-shelf document embeddings~\citep{Tay2022TransformerMA,Wang2022ANC}. 
However, such rule-based methods are usually ad-hoc and do not generalize well. 
In particular, the tokenization results potentially perform well on retrieving documents that have been seen during training, but generalize poorly to new or out-of-distribution documents~\citep{Mehta2022DSIUT,Lee2022ContextualizedGR}.

\header{The proposed method}
To address the above problem, this paper proposes \textsc{GenRet}, a document tokenization learning framework that learns to tokenize a document into semantic docids in a discrete auto-encoding scheme.
Specifically, \textsc{GenRet} consists of a shared sequence-to-sequence-based document tokenization model, a retrieval model, and a document reconstruction model.
In the proposed auto-encoding learning scheme, the tokenization model learns to convert documents to discrete docids, which are subsequently utilized by the reconstruction model to reconstruct the original document.
The generative retrieval model is trained to generate docids in an autoregressive manner for a given query.
The above three models are optimized in an end-to-end fashion to achieve seamless integration.

We further identify two challenges when using auto-encoding to optimize a generative retrieval model: 
(i) docids with an autoregressive nature, and (ii) docids with diversity.
{To address the first challenge} and also to stabilize the training of \textsc{GenRet}, we devise a progressive training scheme.
This training scheme allows for a stable training of the model by fixing optimized prefix docids $z_{<t}$.
To optimize the docids at each step, three proposed losses are utilized:
\begin{enumerate*}[label=(\roman*)]
\item a reconstruction loss for predicting the document using the generated docid, 
\item a commitment loss for committing the docid and to avoid forgetting, and 
\item a retrieval loss for optimizing the retrieval performance end-to-end. 
\end{enumerate*}
{To address the second challenge}, we propose a parameter initialization strategy and a re-assignment of the docid based on a \emph{diverse clustering} technique to increase the diversity of the generated docids. 

\header{Experiments}
We conduct experiments on three well-known document retrieval benchmark datasets:
\begin{enumerate*}[label=(\roman*)]
    \item NQ320K, with a subset of Wikipedia~\citep{Kwiatkowski2019NaturalQA,Tay2022TransformerMA};
    \item {MS~MARCO}, with web pages relevant to a set of search queries~\citep{Campos2016MSMA,Zhou2022UltronAU}; and
    \item BEIR, with heterogeneous retrieval tasks for out-of-distribution evaluation~\citep{Thakur2021BEIRAH}.
\end{enumerate*}
Our experimental results demonstrate that \textsc{GenRet} attains superior retrieval performance against state-of-the-art dense or generative retrieval models.
Experiments on NQ320K show that \textsc{GenRet} establishes the new state-of-the-art on this dataset, achieving +14\% relative improvements on the unseen test set compared to the best baseline method.
Experiments on MS MARCO and six BEIR datasets also shows that \textsc{GenRet} significantly outperforms existing generative methods and achieves competitive results compared to the best dense retrieval model.
Experiments on retrieving new documents, analytical experiments, and efficiency analysis confirm the effectiveness of the proposed model.

\header{Contributions}
In this paper we make the following contributions:
\begin{enumerate*}[label=(\roman*)]
    \item We propose \textsc{GenRet}, a generative retrieval model that represents documents as semantic discrete docids. 
    To the best of our knowledge, this is the first tokenization learning method for document retrieval.
    \item We propose an auto-encoding approach, where the docids generated by our tokenization model are reconstruct by a reconstruction model to ensure the docids capture the semantic information of the document.
    \item We devise a progressive training scheme to model the autoregressive nature of docids and stabilize the training process.
    \item Experimental results demonstrate that \textsc{GenRet} achieves significant improvements, especially on unseen documents, compared to generative retrieval baselines.
\end{enumerate*}

%% file: sections/03-Preliminary.tex
\section{Preliminaries} 
\label{preliminary}

The document retrieval task can be formalized as the process of retrieving a relevant document $d$ for a search query $q$ from a collection of documents $\mathcal{D}$.
Each document, $d \in \mathcal{D}$, is a plain text document consisting of a sequence of tokens, denoted as $d=\{d_1,\ldots,d_{|d|}\}$, where $|d|$ represents the total number of tokens in the document.

Unlike dense retrieval methods, which return the most relevant documents based on the relevance score of each document with respect to a given query $q$, \emph{generative retrieval} models aim to directly generate documents for a given query $q$ using a generative model.

\begin{figure*}[t]
 \centering
\includegraphics[width=0.8\textwidth]{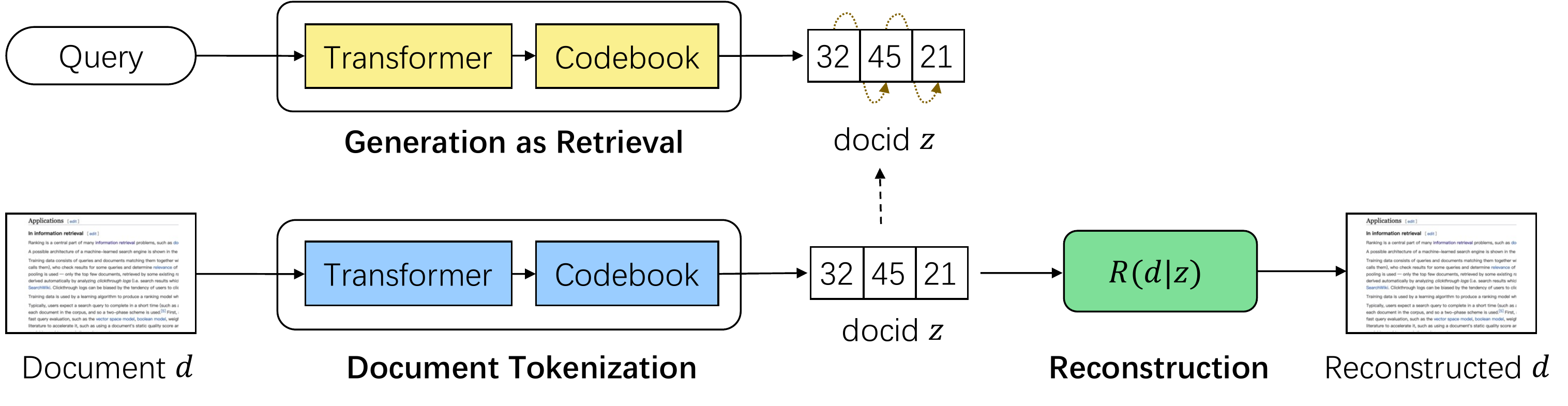} 
\caption{
An overview of the proposed method.
The proposed method utilizes a document tokenization model to convert a given document into a sequence of discrete tokens, referred to as a docid. 
This tokenization process allows for the reconstruction of the original document through a reconstruction model. 
Subsequently, an autoregressive generation model is employed to retrieve documents through the generation of their respective docids.
}
\label{fig:model}
\end{figure*}

\header{Document tokenization}
For \emph{generative retrieval} models, it is usually challenging and computationally inefficient to directly generate original documents of typically long length.
Therefore, most existing approaches rely on the technique named \emph{document tokenization}, which represents a document $d=\{d_1,\ldots,d_{|d|}\}$ as a shorter sequence of discrete tokens (docid) $z=\{z_1,\ldots,z_t,\ldots,z_M\}$, where each token $z_t$ is as a $K$-way categorical variable, with $z_t \in {1,2,\ldots,K}$.

As an alternative sequence of the original document, the tokenized docid $z$ should satisfy the following two properties: 
\begin{enumerate*}[label=(\roman*)]
    \item different documents have short but different docids;
    \item docids capture the semantics of their associated documents as much as possible~\citep{Tay2022TransformerMA}.
\end{enumerate*}
Because $z$ is a sequence of a fixed length and usually shorter than the original document $d$, the model's training and inference can be simplified and more efficient.

As mentioned above, this paper employs a tokenization model $Q\colon d \rightarrow z$ to map $d$ to docid $z$.
More details about $Q$ are provided in Section~\ref{sec:model-architecture}.

\header{Generation as retrieval}
After tokenizing each document to docid $z$, a generative retrieval model $P\colon q \rightarrow z$ learns to retrieve relevant documents by generating a query $q$ to a docid $z$ autoregressively:
\begin{equation}\label{eq:gen-as-ret}
    z = \prod_{t=1}^{M} P(z_t\mid z_{<t}, q),
\end{equation}
where $z_{<t}$ denotes the prefix of $z$ up to time step $t$. 
The model employs a \emph{constrained decoding} technique to ensure that the generated docid $z$ exists in the corpus $\mathcal{D}$~\citep{DeCao2020AutoregressiveER}. 
This is achieved by constructing a prefix tree based on the valid docids in $\mathcal{D}$ and truncating the generation probability of invalid docids to 0.0 during the decoding process. 
The model retrieves multiple documents using beam search.

%% file: sections/04-Method.tex
\section{GenRet}

Conventionally, document tokenization is done by a fixed pre-processing step, such as using the title of a document or the results of hierarchical clustering obtained from BERT~\citep{DeCao2020AutoregressiveER,Tay2022TransformerMA}.
However, it has been observed that such ad-hoc document tokenization methods often fail to capture the complete semantics of a document.
For example, the title of a web page often does not exist or has low relevance to the content of the web page, and the use of clustering-based docids arbitrarily defines the document in discrete space.

In this paper, we propose \textsc{GenRet}, a novel tokenization learning method based on discrete auto-encoding, to learn semantic docid in a fully end-to-end manner.
Figure~\ref{fig:model} gives an overview of the proposed method.
The proposed \textsc{GenRet} comprises three main components: 
(i) a sequence-to-sequence based retrieval model ($P(z\mid q)$), 
(ii) a document tokenization model ($Q(z\mid d)$), and 
(iii) a reconstruction model ($R(d\mid z)$).
The document tokenization model tokenizes a document $d$ into unique discrete variables $z$, and the retrieval model is trained to generate the latent variables $z$ for a given query $q$.
In addition, the reconstruction model is used to re-generate the original document from $z$ to ensure $z$ captures the semantics of the original document as much as possible.

We detail the model architecture of the document tokenization and document retrieval model in Section~\ref{sec:model-architecture}, the devised reconstruction model in Section~\ref{sec:reconstruction-model}, and the model optimization method in Section~\ref{sec:optimization}.

\vspace*{-2mm}
\subsection{Document tokenization and retrieval model}
\label{sec:model-architecture}
Since document tokenization and generative retrieval both aim to map the input text to a discrete docid, we use a shared T5 Transformer architecture for document tokenization and generative retrieval models.
Specifically, given an input text $d$\footnote{We use document $d$ here for the denotation, noting that the computation is the same when $q$ is input.}, 
the T5-based tokenization model encodes $d$ and a prefix of docid $z_{<t}$ and continuously produces latent representation $\mathbf{d}_t$ of $d$ at time step $t$:
\begin{equation}
\label{eq:continuous}
    \mathbf{d}_t = \operatorname{Decoder}(\operatorname{Encoder}(d), z_{<t}) \ \in \mathbb{R}^{D},
\end{equation}
where $D$ denotes the hidden size of the model, $\operatorname{Encoder}(d)$ denotes the output of the $\operatorname{Encoder}$. 

Then, the tokenization model generates a token for each document based on $\mathbf{d}_t$.
At each timestep $t$, we define an external embedding matrix named \emph{codebook} $\mathbf{E}_t \in \mathbb{R}^{K \times D}$, where $K$ is the size of the discrete latent space.
%Here, $K$ can be regarded as the number of semantic space segmentation at timestep $t$.
There are $K$ embedding vectors $\mathbf{e}_{t,j} \in \mathbb{R}^D, j \in [K]$, and each vector $\mathbf{e}_{t,j}$ can be regarded as the centroid of a segmentation.

Based on the \emph{codebook} $\mathbf{E}_t$, the discrete latent variable $z_t$ at timestep $t$ is calculated by a dot-product look-up using the codebook $\mathbf{E}_t$:
\begin{equation}
\label{eq:softmax-e}
    Q(z_t=j\mid z_{<t},d) = \operatorname{Softmax}_j(\mathbf{d}_t \cdot \mathbf{E}_t^\top),
\end{equation}
where $Q(z_t=j\mid z_{<t},d)$ denotes the probability of tokenizing $d$ to a particular value $j \in [K]$ at timestep $t$,
$\operatorname{Softmax}_j$ is a softmax function to output the probability of axis $j$.

Finally, the tokenization model selects the docid that achieves the maximum probability to define the docid $z_t$: 
\begin{equation}
\label{eq:define-z}
    z_t = \argmax_j Q(z_t=j\mid z_{<t},d).
\end{equation}
in which the model selects the id $j$ corresponding to the embedding vector $\mathbf{e}_{t,j}$ with the maximum inner-product with $\mathbf{d}_t$ as the docid $z_t$ at timestep $t$.

The generative retrieval model $P(z\mid q)$ shares the same architecture as $Q(z\mid d)$, while generating $z$ using the input query $q$, as formulated in Eq.~\ref{eq:gen-as-ret}.

\vspace*{-2mm}
\subsection{Document reconstruction model}
\label{sec:reconstruction-model}
The docid generated by the tokenization model $Q$ is required to capture the semantic information of the document.
To this end, we propose an auto-encoding training scheme, where a reconstruction model $R\colon z \rightarrow d$ that predicts $d$ using $z$ is designed to force the tokenization model $Q\colon d \rightarrow z$ to reproduce a docid $z$ that can be reconstructed back-to-the original document.

The input of the reconstruction model is docid $z$, and the output is its associated document $d$.
We first embed $z$ into representation matrix $\mathbf{z}=\{\mathbf{z}_1,\dots,\mathbf{z}_M\} \in \mathbb{R}^{M \times D}$ using the codebook of the tokenization model:
\begin{equation}\label{eq:quant-z}
    \mathbf{z} = \{\mathbf{e}_{1,z_1},\mathbf{e}_{2,z_2},\ldots,\mathbf{e}_{M,z_M}\} \in \mathbb{R}^{M \times D},
\end{equation}
where each $t\in[M]$, $\mathbf{z}_t = \mathbf{e}_{t,z_t} \in \mathbb{R}^D$ is the embedding vector of $z_t$ in the $t$-step codebook $\mathbf{E}_t$.

We then devise a retrieval-based reconstruction model that predicts the target document $d$ by retrieving it from document collection $\mathcal{D}$, based on the inputs $\mathbf{z}$.
The relevance score between the input docid $z$ and the target document $d$ is defined as follows:
\begin{equation}\label{eq:retrieval-rel}
    R(d\mid \mathbf{z}) = \prod_{t=1}^{M} \frac{\exp(\mathbf{z}_t \cdot \operatorname{sg}(\mathbf{d}_t^\top))}{\sum_{d^*\in S(z_{<t})} \exp(\mathbf{z}_t \cdot \operatorname{sg}(\mathbf{d^*}_t^\top))},
\end{equation}
where $S(z_{<t})$ is a sub-collection of $\mathcal{D}$ consisting of documents that have a docid prefix that is the same as $z_{<t}$.
$d^*\in S(z_{<t})$ represents a document from the sub-collection $S(z_{<t})$.
$\mathbf{d}_t$ and $\mathbf{d^*}_t$ are continuous representation of documents $d$ and $d^*$, respectively, as defined in Eq.~\ref{eq:continuous}. 
The operator $\operatorname{sg}(\cdot)$ is the stop gradient operator defined as follows:
\begin{equation}
\operatorname{sg}(x) = 
\begin{cases}
x, &\text{forward pass}\\
0, &\text{backward pass.}
\end{cases}
\end{equation}
Intuitively, $R(d\mid \mathbf{z})$ is designed to retrieve a specific document $d$ from a set of documents $S(z_{<t})$ at each timestep $t$.
The set $S(z_{<t})$ only includes those documents that are assigned the same docid prefix $z_{<t}$ as the target document $d$.
By utilizing this loss function, at each step $t$, the model is facilitated to learn the residual semantics of the documents not captured by the previous docid $z_{<t}$.

\vspace*{-2mm}
\subsection{Model optimization}
\label{sec:optimization}
For the document tokenization model $Q(z\mid d)$, generative retrieval model $P(z\mid q)$ and reconstruction model $R(d\mid z)$, jointly optimizing these three models using auto-encoding is challenging for the following two reasons:
\begin{itemize}[leftmargin=*]
    \item \textbf{Learning docids in an autoregressive fashion}. That is: 
    \begin{enumerate*}[label=(\roman*)]
        \item The prediction of the $z_t$ at time $t$ relies on previously predicted docids $z_{<t}$, which is often under-optimized at the beginning and rapidly changes during training, making it difficult to reach convergence.
        \item Simultaneously optimizing $z$ makes it challenging to guarantee a unique docid assignment.
    \end{enumerate*}
    To stabilize the training of \textsc{GenRet}, we devise a \emph{progressive training scheme} (see Section~\ref{sec:auto-encoding}).

    \item \textbf{Generating docids with diversity}. Optimizing the model using auto-encoding often leads to unbalanced docid assignment: a few major docids are assigned to a large number of documents and most other docids are rarely assigned.
    Such a sub-optimal distribution of docids affects the model distinguishability, which in turns triggers length increments of docids in order to distinguish conflicting documents.
    We introduce two \emph{diverse clustering} techniques to ensure docid diversity (see Section~\ref{sec:diverse-cluster}).
\end{itemize}

\begin{figure}[t]
 \centering
\includegraphics[width=1\columnwidth]{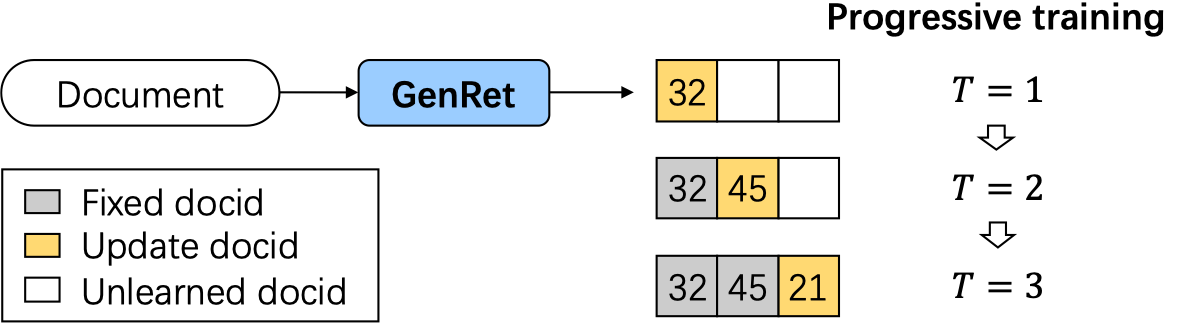} 
\caption{
Progressive training scheme.
$z_t$ (docid at timestep $t$) is optimized at the $t$-th training step, while $z_{<t}$ (docids before timestep $t$) are kept fixed.
}
\label{fig:training}
\end{figure}

\subsubsection{Progressive training scheme}
\label{sec:auto-encoding}

To optimize each of the three models listed above in an autoregressive manner, we propose a progressive auto-encoding learning scheme, as illustrated in Figure~\ref{fig:training}.
The whole learning scheme contains $M$ learning steps with respect to the final docid in $M$-token.
And the docid $z_T$ at step $T \in [M]$ is learned and optimized at the corresponding learning step.
Besides, at each step $T \in [M]$, the docid $z_T$ and the model parameters associated with $z_T$ generation are updated, while previously produced docids $z_{<T}$ and other parameters are kept fixed.
By progressively performing the above process, we can finally optimize and learn our models.

At each optimization step, say the $T$-step, we devise the learning objective for document tokenization consisting of three loss functions detailed below.

\header{Reconstruction loss}
We utilize the reconstruction model $R(d\mid z)$ as an auxiliary model to learn to optimize the docid generation, whose main goal is capturing as much semantics in the docid as possible.
Therefore, we define a reconstruction loss function of step $T$ as follows:
\begin{equation}
\begin{split}
\mathcal{L}_{\text{Rec}} &= -\log R(d\mid \hat{\mathbf{z}}_{\le T})
\\
\hat{\mathbf{z}}_{\le T} &= \{\operatorname{sg}(\mathbf{z}_1),\dots,\operatorname{sg}(\mathbf{z}_{T-1}),\mathbf{z}_T\} \ \in \mathbb{R}^{T \times D}
\\
\forall t \in [T]: \ \mathbf{z}_t &= \mathbf{e}_{t, j^*} \in \mathbb{R}^D, \ j^* = \argmax_j Q(z_t=j\mid z_{<t},d),
\end{split}
\end{equation}
where $\hat{\mathbf{z}}_{\le T}$ is the first $T$ representations of the $z$, and only the variable $\mathbf{z}_T$ is optimized in step $T$.
$Q(z_t=j\mid z_{<t},d)$ is defined in Eq.~\ref{eq:softmax-e}.
And the document tokenization model $Q$ can therefore be optimized when minimizing $\mathcal{L}_{\text{Rec}}$.

Of note, since the computation involves a non-differentiable operation -- $\argmax(\cdot)$, we apply straight-through gradient estimation to back-propagate the gradient from reconstruction loss~\citep{Oord2017NeuralDR,Zhan2021LearningDR}.
Specifically, the gradients to document representation $\mathbf{d}_T$ are defined as 
$\frac{\partial \mathcal{L}_{\text{Rec}}}{\partial \mathbf{d}_T} \coloneqq 
\frac{\partial \mathcal{L}_{\text{Rec}}}{\partial \mathbf{z}_T}$.
And the gradients to the \emph{codebook} embedding $\mathbf{e}_{T,j}$ are defined as 
$\frac{\partial \mathcal{L}_{\text{Rec}}}{\partial \mathbf{e}_{T,j}} \coloneqq 
1_{z_T=j}
\frac{\partial \mathcal{L}_{\text{Rec}}}{\partial \mathbf{z}_T}$.

\header{Commitment loss}
In addition, to make sure the predicted docid commits to an embedding and to avoid models forgetting previous docid $z_{<t}$, we add a commitment loss as follows:
\begin{equation}
    \mathcal{L}_{\text{Com}} = -\sum_{t=1}^{T}\log Q(z_t\mid z_{<t}, d).
\end{equation}

\header{Retrieval loss}
For the generative retrieval model $P$, we jointly learn it together with the document tokenization model $Q$, where $P$ learns to generate the docids of relevant documents $d$ given a query $q$.
Specifically, suppose $(q,d)$ are a query and relevant document pair; we define the learning objective of retrieval model $P$ as:
\begin{equation}
    \mathcal{L}_{\text{Ret}} = -\log \frac{\exp(\mathbf{q}_T \cdot \mathbf{d}_T)}{\sum_{d^- \sim B} \exp(\mathbf{q}_T \cdot \mathbf{d^-}_T)} - \sum_{t=1}^{T}\log P(z_t\mid z_{<t}, q),
\end{equation}
% \zhc{$z_t=?$}
where the first term is a ranking-oriented loss enhancing the model using $(q,d)$ pair;
$d^-$ is an in-batch negative document from the same training mini-batch $B$; $\mathbf{q}_T$ and $\mathbf{d}_T$ denote the representation of $q$ and $d$ at timestep $T$.
The second term is the cross-entropy loss for generating docid $z$ based on query $q$.

The final loss we use at step-$T$ is the sum of reconstruction loss, commitment loss, and retrieval loss:
\begin{equation}
    \mathcal{L} = \mathcal{L}_{\text{Rec}} + \mathcal{L}_{\text{Com}} + \mathcal{L}_{\text{Ret}}.
\end{equation}

\subsubsection{Diverse clustering technique}
\label{sec:diverse-cluster}
To ensure diversity of generated docids, we adopt two diverse clustering techniques--codebook initialization and docid re-assignment at each progressive training step, where codebook initialization mainly aims to increase the balance of semantic space segmentation, and the docid re-assignment mainly aims to increase the balance of docid assignments.

\header{Codebook initialization} 
In order to initialize the codebook for our model, we first warm-up the model by passing the continuous representation $\mathbf{d}_T$ to the reconstruction model instead of the docid representation $\mathbf{z}_T$ as defined in Eq.~\ref{eq:quant-z}. 
During this warm-up phase, we optimize the model using the reconstruction loss $\mathcal{L}_{\text{Rec}}$ and commitment loss $\mathcal{L}_{\text{Com}}$. 
Next, we collect the continuous representations $\mathbf{d}_T$ of all documents in $\mathcal{D}$, and cluster them into $K$ groups. 
The centroids of these clusters are then used as the initialized codebook $\mathbf{E}_T$. 
To balance the initialized docid distribution, we utilize a diverse constrained clustering algorithm, \emph{Constrained K-Means}, which modifies the cluster assignment step (E in EM) by formulating it as a minimum cost flow (MCF) linear network optimization problem~\citep{Bennett2000ConstrainedKC}.

\header{Docid re-assignment} 
In order to assign docids to a batch of documents, we modify the dot-product look-up results in Eq.~\ref{eq:softmax-e} by ensuring that the docid for different documents in the batch are distinct~\citep{Caron2020UnsupervisedLO,Zhan2021LearningDR}. 
Specifically, let $\mathbf{D}_t=\{\mathbf{d}_t^{(1)}, \ldots, \mathbf{d}_t^{(B)}\} \in \mathbb{R}^{B \times D}$ denote the continuous representation of a batch of documents with batch size of $B$.
The dot-product results are represented by $\mathbf{H} = \mathbf{D}_t \cdot \mathbf{E}_t^\top \in \mathbb{R}^{B \times K}$.
To obtain distinct docids, we calculate an alternative
$\mathbf{H}^* = \operatorname{Diag}(\mathbf{u}) \exp(\frac{\mathbf{H}}{\epsilon}) \operatorname{Diag}(\mathbf{v})$,
where $\mathbf{u}$ and $\mathbf{v}$ are re-normalization vectors in $\mathbb{R}^K$ and $\mathbb{R}^B$, respectively. 
The re-normalization vectors are computed via the iterative Sinkhorn-Knopp algorithm~\citep{Cuturi2013SinkhornDL}.
Finally, $\mathbf{H}^*$ is used instead of $\mathbf{H}$ in the $\operatorname{Softmax}$ (Eq.~\ref{eq:softmax-e}) and $\argmax$ (Eq.~\ref{eq:define-z}) operations to obtain the docid $z_t$.

%% file: sections/05-Experiment.tex
\section{Experimental Setup}

\subsection{Datasets}
We conduct experiments on three well-known document retrieval datasets: NQ~\citep{Kwiatkowski2019NaturalQA}, MS MARCO~\citep{Campos2016MSMA}, and BEIR~\citep{Thakur2021BEIRAH}.

\header{NQ320K}
NQ320K is a popular dataset for evaluating generative retrieval models~\citep{Tay2022TransformerMA,Wang2022ANC}. It is based on the Natural Questions (NQ) dataset proposed by Google~\citep{Kwiatkowski2019NaturalQA}. 
NQ320k consists of 320k query-document pairs, where the documents are gathered from Wikipedia pages, and the queries are natural language questions.
We follow the evaluation setup in NCI~\citep{Wang2022ANC} and further split the test set into two subsets: 
\emph{seen test}, in which the annotated target documents of the queries are included in the training set; and 
\emph{unseen test}, in which no labeled document is included in the training set.

\header{MS MARCO}
MS MARCO is a collection of queries and web pages from Bing search. 
Akin to NQ320k and following \citep{Zhou2022UltronAU}, we sample a subset of documents from the labeled documents, and use their corresponding queries for training. 
We evaluate the models on the queries of the MS MARCO dev set and retrieval on the sampled document subset.

\header{BEIR}
BEIR is a collection of datasets for heterogeneous retrieval tasks.
In this paper, we evaluate the models on 6 BEIR datasets, which include distinct retrieval tasks and document collections from NQ and MS MARCO:
(i) BEIR-Arg retrieves a counterargument to an argument;
(ii) BEIR-Covid  retrieves scientific articles about the COVID-19 pandemic;
(iii) BEIR-NFC retrieves medical documents from PubMed;
(iv) BEIR-SciFact retrieves scientific papers for fact-checking;
(v) BEIR-SciDocs retrieves citations for scientific papers;
(vi) BEIR-FiQA retrieves financial documents.

We summarize the statistics of above datasets in Table~\ref{table:data}.

\input{tables/data.tex}

% \vspace*{-2mm}
\subsection{Evaluation metrics}
On NQ320K, we use Recall@\{1,10,100\} and Mean Reciprocal Rank (MRR)@100 as evaluation metrics, following~\citep{Wang2022ANC}.
On MS MARCO, we use Recall@\{1, 10, 100\} and MRR@10 as evaluation metrics, following~\citep{Zhou2022UltronAU}.
On BEIR, we use nDCG@10 as the main metrics and calculate the average nDCG@10 values across multiple downstream sub-datasets as overall metrics.

% \vspace*{-1mm}
\subsection{Baselines}
We consider three types of baselines: sparse retrieval methods, dense retrieval methods, and generative retrieval methods.
%

% \noindent
The sparse retrieval baselines are as follows:
\begin{itemize*}[leftmargin=*]
    \item \textbf{BM25}, uses the tf-idf feature to measure term weights; we use the implementation from \url{http://pyserini.io/}.
    \item \textbf{DocT5Query}, expands a document with possible queries predicted by a finetuned T5 with this document as the input.
\end{itemize*}

The dense retrieval baselines are as follows:
\begin{itemize*}[leftmargin=*]
    \item \textbf{DPR}~\citep{Karpukhin2020DensePR}, a dual-encoder model using the representation of the \code{[CLS]} token of BERT.
    \item \textbf{ANCE}~\citep{Xiong2020ApproximateNN}, an asynchronously updated ANN indexer is utilized to mine hard negatives for training a RoBERTa-based dual-encoder model.
    \item \textbf{Sentence-T5}~\citep{Ni2021SentenceT5SS}, a dual-encoder model that uses T5 to produce continuous sentence embeddings. We reproduce Sentence-T5 (ST5 for short) on our datasets, the model is based on T5-Base EncDec model and is trained with in-batch negatives.
    \item \textbf{GTR}~\citep{Ni2021LargeDE}, a state-of-the-art dense retrieval model that pre-trains sentence-T5 on billions of paired data using contrastive learning.
    \item \textbf{Contriever}~\citep{Izacard2021UnsupervisedDI}, a dual-encoder model pre-trained using unsupervised contrastive learning with independent cropping and inverse cloze task.
\end{itemize*}

And the generative retrieval baselines are as follows:
\begin{itemize*}[leftmargin=*,nosep]
    \item \textbf{GENRE}~\citep{DeCao2020AutoregressiveER}, an autoregressive retrieval model that generates the document's title. The original GENRE is trained on the KILT dataset~\citep{Petroni2020KILTAB} using BART, and we reproduce GENRE on our datasets using T5 for a fair comparison. For datasets without title, we use the first 32 tokens of the document as pseudo-title.
    \item \textbf{DSI}~\citep{Tay2022TransformerMA}, which represents documents using hierarchical K-means clustering results, and indexes documents using the first 32 tokens as pseudo-queries. As the original code is not open source, we reproduce DSI using T5-base and the docids of NCI~\citep{Wang2022ANC}.
    \item \textbf{SEAL}~\citep{Bevilacqua2022AutoregressiveSE} uses arbitrary n-grams in documents as docids, and retrieves documents under the constraint of a pre-built FM-indexer. We refer to the results reported by \citet{Wang2022ANC}.    
    \item \textbf{CGR-Contra}~\citep{Lee2022ContextualizedGR}, a title generation model with a contextualized vocabulary embedding and a contrastive learning loss.    
    \item \textbf{DSI-QG}~\citep{Zhuang2022BridgingTG}, uses a query generation model to augment the document collection. We reproduce the DSI-QG results using T5 and our dataset.    
    \item \textbf{NCI}~\citep{Wang2022ANC}, uses a prefix-aware weight-adaptive decoder and various query generation strategies, including DocAsQuery and DocT5Query. In particular, NCI augments training data by generating 15 queries for each document.    
    \item \textbf{Ultron}~\citep{Zhou2022UltronAU}, uses a three-stage training pipeline and represents the document as three types of identifiers, including URL, PQ, and Atomic.
\end{itemize*}

We highlight three of our reproduced baselines that constitute a fair comparison with the proposed method, all of which use the T5 model and experimental setup, but they differ model outputs: 
(i) Sentence-T5 outputs continuous vectors, 
(ii) GENRE outputs document titles, 
(iii) DSI-QG outputs clustering ID, while
\textsc{GenRet} outputs docids learned using the proposed tokenization method.

\subsection{Implementation details}
\textbf{Hyper-parameters.}
In our experiments, we utilize the T5-Base model~\citep{Raffel2019ExploringTL} as the base Transformer and initialize a new codebook embedding $\mathbf{E}_t$ for each time step.
We set the number of clusters to be $K=512$ for all datasets, with the length of the docid $M$ being dependent on the number of documents present. For datasets containing a larger number of candidate documents, a larger value of $M$ is set to ensure that all documents are assigned unique document ids.
In the docid re-assignment, the hyper-parameter $\epsilon$ is set to $1.0$, and the Sinkhorn-Knopp algorithm is executed for $100$ iterations.

\header{Indexing with query generation}
Following previous work~\citep{Zhuang2022BridgingTG,Wang2022ANC,Wang2021GPLGP}, we use query generation models to generate synthetic 
(query, document) pairs for data augmentation.
Specifically, we use the pre-trained query generation model from DocT5Query~\citep{Cheriton2019FromDT} to augment the NQ and MS MARCO datasets. 
In query generation, we use nucleus sampling with parameters $p=0.8, t=0.8$ and generate five queries for each document in the collection.
For the BEIR datasets, we use the queries generated by GPL~\citep{Wang2021GPLGP}, which can be downloaded from their website.\footnote{\url{https://public.ukp.informatik.tu-darmstadt.de/kwang/gpl/generated-data/beir/}}
GPL uses a DocT5Query~\citep{Cheriton2019FromDT} generator trained on MS MARCO to generate about 250K queries for each BEIR dataset.

\header{Training and inference}
The proposed models and the reproduced baselines are implemented with PyTorch 1.7.1 and HuggingFace transformers 4.22.2.
We optimize the model using AdamW and set the learning rate to $5e-4$.
The batch size is $256$, and the model is optimized for up to 500k steps for each timestep.
In progressive training, we first warm up the model for 5K steps and then initialize the codebook using the clustering centroids as mentioned in Section~\ref{sec:auto-encoding}.
We use constrained clustering\footnote{\url{https://github.com/joshlk/k-means-constrained}} to obtain diverse clustering results.
During inference, we use beam search with constrained decoding~\citep{DeCao2020AutoregressiveER} and a beam size of $100$.

%% file: tables/data.tex
\begin{table}[!t]
\centering
\setlength\tabcolsep{5pt}
\caption{Statistics of datasets used in our experiments. 
The three values split by / on \# Test queries denote the number of queries in the full, seen subset, and unseen subset, respectively.
In BEIR, all queries in the test set are unseen.
}
\label{table:data}
\begin{tabular}{l ccc}
\toprule
Dataset & \# Docs & \# Test queries & \# Train pairs\\
\midrule
NQ320K & 109,739 & 7,830 / 6,075 / 1,755 & 307,373 \\
MS MARCO & 323,569 & 5,187 / \phantom{0,}807 / 4,380 & 366,235 \\
\midrule
BEIR-Arg &  \phantom{00}8,674 & 1,406 & - \\
BEIR-Covid & 171,332 & \phantom{00}50 & - \\
BEIR-NFC & \phantom{00}3,633 & \phantom{0}323 & - \\
BEIR-SciFact & \phantom{00}5,183 & \phantom{0}300 & - \\
BEIR-SciDocs &  \phantom{0}25,657 & 1,000 & - \\
BEIR-FiQA & \phantom{0}57,638 & \phantom{0}648 & -\\

\bottomrule
\vspace*{1mm}
\end{tabular}
\end{table}

%% file: sections/06-Results.tex
\input{tables/nq320k.tex}
\input{tables/msmarco.tex}
\input{tables/beir-lite.tex}

% \vspace*{-2mm}
\section{Experimental results}

\subsection{Main results}

\textbf{Results on NQ320K.}
In Table~\ref{table:nq320k}, we list the results on NQ320K.
\textsc{GenRet} outperforms both the strong pre-trained dense retrieval model, GTR, and the previous best generative retrieval method, NCI, thereby establishing a new state-of-the-art on the NQ320K dataset.
Furthermore, our results reveal that existing generative retrieval methods perform well on the seen test but lag behind dense retrieval methods on the unseen test.
For example, NCI obtains an MRR@100 of $76.8$ on the seen test, which is higher than the MRR@100 of $65.3$ obtained by GTR-Base. 
However, on unseen test data, NCI performs worse than GTR-Base. In contrast, \textsc{GenRet} performs well on both seen and unseen test data.
This result highlights the ability of \textsc{GenRet} to combine the advantages of both dense and generative retrieval by learning discrete docids with semantics through end-to-end optimization.

\header{Results on MS MARCO}
Table~\ref{table:msmarco} presents the results on the MS MARCO dataset.
\textsc{GenRet} significantly outperforms previous generative retrieval methods and achieves comparable results with the state-of-the-art dense retrieval method GTR.
Furthermore, previous generative retrieval methods (e.g., GENRE, Ultron) utilizing metadata such as the title and URL, while exhibiting decent performance on the NQ320K dataset, underperform in comparison to previous-best dense retrieval (GTR) and sparse retrieval (DocT5Query) methods on the MS MARCO dataset. 
This can likely because that the NQ320K dataset retrieves Wikipedia documents, where metadata like the title effectively capture the semantics of the document. 
In the case of the MS MARCO dataset, which is a web search dataset, the metadata often does not adequately characterize the documents, resulting in a decline in performance of the generative retrieval model. 
In contrast, \textsc{GenRet} learns to generate semantic docids that effectively enhance the generative retrieval model.

\header{Results on BEIR}
Table~\ref{table:beir} lists the results of the baselines and \textsc{GenRet} on six datasets of BEIR.
These datasets represent a diverse range of information retrieval scenarios.
On average, \textsc{GenRet} outperforms strong baselines including BM25 and GTR-Base, and achieves competitive results compared to state-of-the-art sparse and dense retrieval methods.
In comparison to the ST5 GPL method that utilizes the same training data and backbone T5 model, \textsc{GenRet} achieves better results.
Additionally, \textsc{GenRet} demonstrates a significant improvement over the previous generative retrieval model GENRE that utilizes titles as docids.
Furthermore, GENRE performs poorly on some datasets, such as BEIR-Covid and BEIR-SciDocs.
This may be because the titles of the documents in these datasets do not adequately capture their semantic content.

\input{tables/ablation-id.tex}

\vspace*{-2mm}
\subsection{Performance on retrieving new documents}
In this experiment, we investigate the impact of various document tokenization techniques on the ability of generative retrieval models to retrieve new documents.
The generative models with different tokenization methods are trained on NQ320K data, excluding unseen documents, and are evaluated on NQ320K Unseen test set and BEIR-\{Arg, NFC, SciDocs\} datasets. 
For the baseline methods, which use rule-based document tokenization methods, the docids are generated for the target document collection using their respective tokenization techniques.
In contrast, our proposed method uses a tokenization model to tokenize the documents in the target collection, producing the docids. 
However, our method may result in duplicate docids. 
In such cases, all corresponding documents are retrieved and shuffled in an arbitrary order.
The results of this evaluation are summarized in Table~\ref{table:aba-id}.

Document tokenization methods that do not consider the semantic information of the documents, such as Naive String and Atomic, are ineffective in retrieving new documents without model updating.
Methods that consider the semantic information of the documents, such as those based on title or BERT clustering, show some improvement.
Our proposed document tokenization method significantly improves over these existing rule-based document tokenization methods.
For instance, when the model trained on NQ -- a factoid QA data based on Wikipedia documents -- is applied to a distinct retrieval task on a different document collection, BEIR-SciDocs, a citation retrieval task on a collection of scientific articles, our proposed document tokenization model still showed promising results with an nDCG@10 of $12.3$, which is comparable to those models trained on the target document collection.
This suggests that our proposed method effectively encodes the semantic information of documents in the docid and leads to a better fit between the docid and the generative retrieval model.

\begin{figure}[t]
 \centering
\includegraphics[width=1\columnwidth]{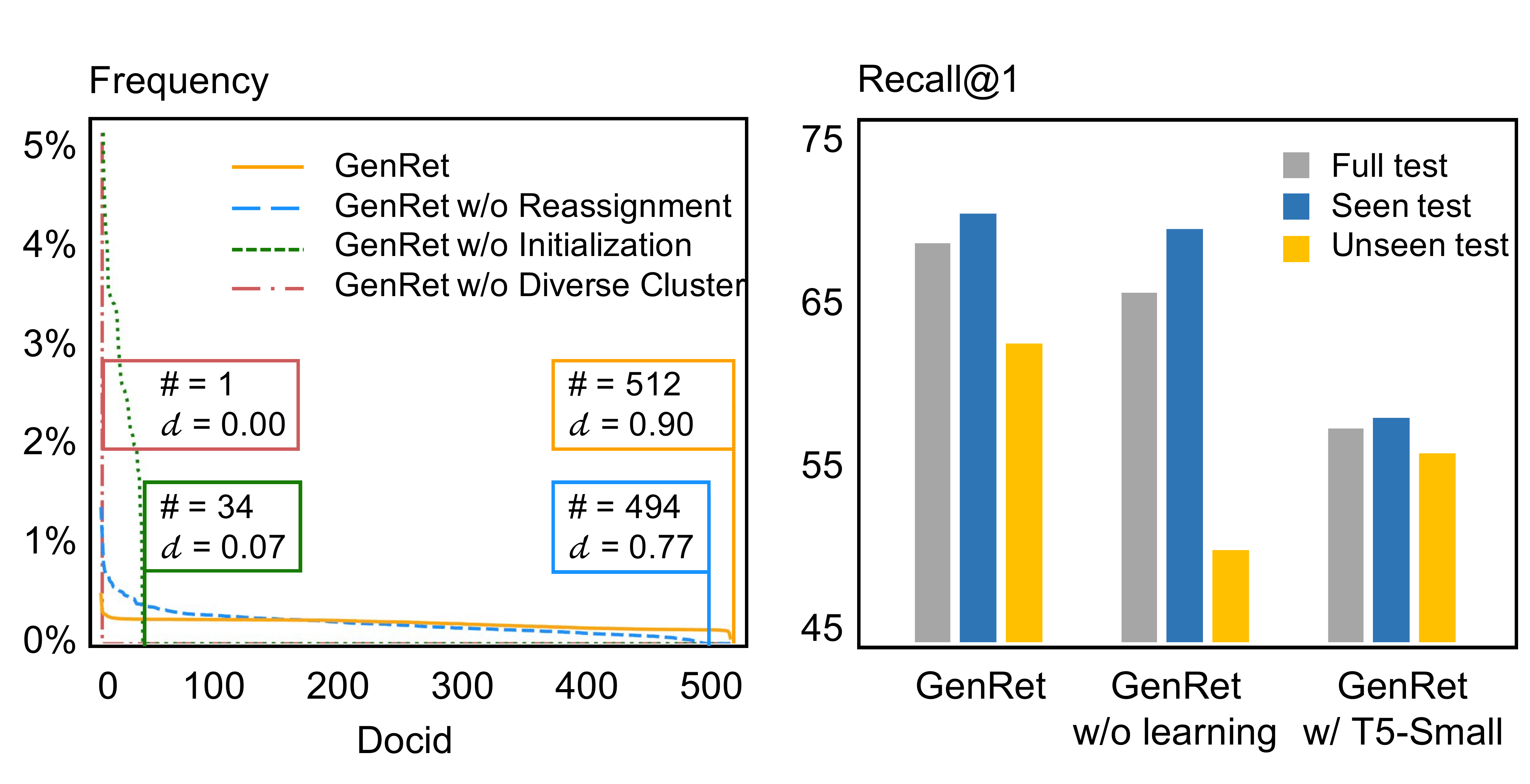} 
\caption{Left: Docid distribution on NQ320K. The id are sorted by the assigned frequency. Right: Ablation study on NQ320K.
}
\label{fig:freq}
\end{figure}

\vspace*{-2mm}
\subsection{Analytical experiments}
We further conduct analytical experiments to study the effectiveness of the proposed method.

In Figure~\ref{fig:freq} (left), we plot the frequencies of docids at the first timestep of various learning methods.
We label each method using a box with a docid and a diversity metric $\mathit{d}$, which is calculated by:
$\mathit{d} = 1 - \frac{1}{2n}\sum_{j=1}^K \left|n_j - n_u\right|$,
where $\left|\cdot\right|$ represents the absolute value, $n$ denotes the total number of documents, $n_j$ denotes the number of documents that have a docid ${}=j$, and $n_u=\frac{n}{K}$ is the expected number of documents per docid under the uniform distribution.

The results demonstrate the superiority of \textsc{GenRet} (represented by the yellow line) in terms of distribution uniformity. 
It uses all the potential docid $k=512$ and achieves the highest diversity metric with a value of $\mathit{d}=0.90$. 
The method without docid reassignment also yields a relatively balanced distribution, with a diversity metric of $\mathit{d}=0.77$. 
However, the distribution of the method without diverse codebook initialization is highly uneven, which can be attributed to the fact that most of the randomly initialized codebook embeddings are not selected by the model during the initial training phase, leading to their lack of update and further selection in subsequent training.
Additionally, the models without diverse clustering
%i.e., those that removed both codebook initialization and docid reassignment, 
tend to converge to a trivial solution where all documents are assigned the same docid.

In Figure~\ref{fig:freq} (right), the results of two ablated variants are presented. 
First, \textsc{GenRet} w/o learning is a generative model that has been trained directly using the final output docid from \textsc{GenRet}, without utilizing the proposed learning scheme. 
Its retrieval performance is comparable to that of \textsc{GenRet} on seen test data; however, it is significantly lower on unseen test data. 
The proposed progressive auto-encoding scheme is crucial for the model to capture the semantic information of documents, rather than just the well-defined discrete docid.
Secondly, \textsc{GenRet} w/ T5-Small uses a small model, and its performance is inferior to that of \textsc{GenRet} using T5-Base. 
However, the gap between the performance on seen and unseen test data is smaller, which could be attributed to the limited fitting capacity of the small model.

\input{tables/speed.tex}

\vspace*{-2mm}
\subsection{Efficiency analysis}
In Table~\ref{table:speed}, we compare \textsc{GenRet} with baseline models on MS MARCO (323,569 documents) in terms of memory footprint, offline indexing time (not including the time for neural network training), and online retrieval latency for different Top-K values.
We have four observations:
(i) The memory footprint of generative retrieval models (GENRE, DSI-QG, and the proposed model) is smaller than of dense and sparse retrieval methods.
The memory footprint of generative retrieval models is only dependent on the model parameters, whereas dense and sparse retrieval methods require additional storage space for document embeddings, which increases linearly with the size of the document collection.
(ii) DSI and \textsc{GenRet} take a longer time for offline indexing, as DSI involves encoding and clustering documents using BERT, while \textsc{GenRet} requires tokenizing documents using a tokenization model.
Dense retrieval's offline time consumption comes from document encoding; GENRE uses titles hence no offline computation.
(iii) The online retrieval latency of the generative retrieval model is associated with the beam size (i.e., Top-K) and the length of the docid.  \textsc{GenRet} utilizes diverse clustering to generate a shorter docid, resulting in improved online retrieval speed compared to DSI and GENRE.

\input{tables/case.tex}

\vspace*{-2mm}
\subsection{Case study}
\label{sec:case-study}
Table~\ref{table:case} shows an example of outputs of GENRE, NCI, and \textsc{GenRet} for the query ``\emph{what state courts can order a new trial}'' and its corresponding document in NQ320K.
The results show that \textsc{GenRet}, unlike the baselines, successfully returns the docid of the target document.
We highlight words in the target document based on their attention activation in \textsc{GenRet} at different time steps $t$.
The yellow color indicates words that received higher attention at $t=1$, while gray indicates words that received higher attention at $t=2$.
The example shows that the model focuses on different words at different time steps.
\textsc{GenRet} gives more attention to words related to the topic, such as \emph{Appellate}, in $t=1$, and more attention to words related to the country, such as \emph{United States}, in $t=2$.

%% file: tables/nq320k.tex
\begin{table*}[!t]
\centering
\setlength\tabcolsep{5pt}
\caption{Results on Natural Questions (NQ320K).
The results of the methods marked with $^\dagger$ are from our own re-implementation, others are from their official implementation.
Methods with $^\spadesuit$ use additional annotated document retrieval data during training.
* and ** indicate significant improvements over previous-best generative retrieval baselines with p-value $< 0.05$ and p-value $< 0.01$, respectively.
$\natural$ and $\sharp$ indicate significant improvements over previous-best dense retrieval baselines with p-value $< 0.05$ and p-value $< 0.01$, respectively.
The best results for each metric are indicated in boldface.
}
\label{table:nq320k}

\begin{tabular}{@{}l cccc  cccc  cccc @{}}
\toprule
& \multicolumn{4}{c}{\textbf{Full test}} 
& \multicolumn{4}{c}{\textbf{Seen test}} 
& \multicolumn{4}{c}{\textbf{Unseen test}} 
\\
\cmidrule(lr){2-5} \cmidrule(lr){6-9} \cmidrule(lr){10-13} 
\textbf{Method}
& R@1 & R@10 & R@100 & MRR@100 & R@1 & R@10 & R@100 & MRR@100 & R@1 & R@10 & R@100 & MRR@100\\

\midrule
\multicolumn{6}{@{}l}{\emph{Sparse retrieval}}\\

BM25~\citep{Robertson2009ThePR}
& 29.7 & 60.3 & 82.1 & 40.2
& 29.1 & 59.8 & 82.4 & 39.5
& 32.3 & 61.9 & 81.2 & 42.7
\\

DocT5Query~\citep{Cheriton2019FromDT}
& 38.0 & 69.3 & 86.1 & 48.9
& 35.1 & 68.3 & 86.4 & 46.7
& 48.5 & 72.9 & 85.0 & 57.0
\\

\midrule
\multicolumn{6}{@{}l}{\emph{Dense retrieval}}\\
DPR~\citep{Karpukhin2020DensePR}
& 50.2 & 77.7 & 90.9 & 59.9
& 50.2 & 78.7 & 91.6 & 60.2
& 50.0 & 74.2 & 88.7 & 58.8
\\

ANCE~\citep{Xiong2020ApproximateNN}
& 50.2 & 78.5 & 91.4 & 60.2
& 49.7 & 79.2 & 92.3 & 60.1
& 52.0 & 75.9 & 88.0 & 60.5
\\

Sentence-T5$^\dagger$~\citep{Ni2021SentenceT5SS}
& 53.6 & 83.0 & 93.8 & 64.1
& 53.4 & 83.9 & 94.7 & 63.8
& 56.5 & 79.5 & 90.7 & 64.9
\\

GTR-Base$^\spadesuit$~\citep{Ni2021LargeDE}
& 56.0 & 84.4 & 93.7 & 66.2
& 54.4 & 84.7 & 94.2 & 65.3
& 61.9 & 83.2 & 92.1 & 69.6
\\

\midrule

\multicolumn{6}{@{}l}{\emph{Generative retrieval}}\\
% GENRE\\
GENRE$^\dagger$~\citep{DeCao2020AutoregressiveER}
& 55.2 & 67.3  & 75.4  & 59.9
& 69.5 & 83.7 & 90.4  & 75.0
& \phantom{0}6.0 & 10.4  & 23.4 & \phantom{0}7.8
\\

DSI$^\dagger$~\citep{Tay2022TransformerMA}
& 55.2 & 67.4 & 78.0  & 59.6
& 69.7 & 83.6 & 90.5 & 74.7
& \phantom{0}1.3 & \phantom{0}7.2 & 31.5  & \phantom{0}3.5
\\

SEAL~\citep{Bevilacqua2022AutoregressiveSE}
& 59.9 & 81.2 & 90.9 & 67.7
& - & - & - & - 
& - & - & - & - 
\\

CGR-Contra~\citep{Lee2022ContextualizedGR}
& 63.4 & 81.1 & - & -
& - & - & - & - 
& - & - & - & - 
\\

DSI-QG$^\dagger$~\citep{Zhuang2022BridgingTG}
& 63.1 & 80.7 & 88.0 & 69.5 
& 68.0 & 85.0 & 91.4 & 74.3
& 45.9 & 65.8 & 76.3 & 52.8
\\

NCI~\citep{Wang2022ANC}
& 66.4 & 85.7 & 92.4 & 73.6
& 69.8 & 88.5 & 94.6 & 76.8
& 54.5 & 75.9 & 84.8 & 62.4
\\

\textbf{Ours}
& \textbf{68.1}\rlap{$^{*\sharp}$} & \textbf{88.8}\rlap{$^{*\natural}$} & \textbf{95.2}\rlap{$^{*}$} & \textbf{75.9}\rlap{$^{*\natural}$}
& \textbf{70.2}\rlap{$^{\sharp}$} & \textbf{90.3}\rlap{$^{\sharp}$} & \textbf{96.0}\rlap{$^{\natural}$} & \textbf{77.7}\rlap{$^{\sharp}$} 
& \textbf{62.5}\rlap{$^{**}$} & \textbf{83.6}\rlap{$^{**}$} & \textbf{92.5}\rlap{$^{**}$} & \textbf{70.4}\rlap{$^{**}$}
\\

\bottomrule
\end{tabular}
\vspace*{-2mm}
\end{table*}

%% file: tables/msmarco.tex
\begin{table}[!t]
\centering
\setlength\tabcolsep{8pt}
\caption{\textbf{Results on MS MARCO.} 
The results of the methods marked with $^\dagger$ are from our own re-implementation. 
Methods with $^\spadesuit$ use additional annotated retrieval data for training.
*/** indicates significant improvements over previous generative retrieval baselines with p-value $< 0.05/0.01$.
The best results for each metric are indicated in boldface.
}
\label{table:msmarco}

\begin{tabular}{@{}l cccc @{}}

\toprule
Method
& R@1 & R@10 & R@100 & MRR@10 \\

\midrule
\multicolumn{4}{@{}l}{\emph{Sparse retrieval}}\\

BM25~\citep{Robertson2009ThePR}
& 39.1 & 69.1 & 86.2 & 48.6
\\

DocT5Query~\citep{Cheriton2019FromDT}
& 46.7 & 76.5 & 90.4 & 56.2
\\

\midrule
\multicolumn{4}{@{}l}{\emph{Dense retrieval}}\\

ANCE~\citep{Xiong2020ApproximateNN}
& 45.6 & 75.7 & 89.6 & 55.6
\\

Sentence-T5$^\dagger$~\citep{Ni2021SentenceT5SS}
& 41.8 & 75.4 & 91.2 & 52.8
\\

GTR-Base$^\spadesuit$~\citep{Ni2021LargeDE}
& 46.2 & 79.3 & \textbf{93.8} & 57.6
\\

\midrule

\multicolumn{4}{@{}l}{\emph{Generative retrieval}}\\
GENRE$^\dagger$~\citep{DeCao2020AutoregressiveER}
& 35.6 & 57.6 & 79.1 & 42.3
\\

Ultron-URL~\citep{Zhou2022UltronAU}
& 29.6 &  67.8 & - & 40.0
\\

Ultron-PQ~\citep{Zhou2022UltronAU}
& 31.6 &  73.1 & - & 45.4
\\

Ultron-Atomic~\citep{Zhou2022UltronAU}
& 32.8 &  74.1 & - & 46.9
\\

\textbf{Ours}
& \textbf{47.9}\rlap{$^{**}$} & \textbf{79.8}\rlap{$^{**}$} & 91.6\rlap{$^{**}$} & \textbf{58.1}\rlap{$^{**}$}
\\

\bottomrule
\end{tabular}
\end{table}

%% file: tables/beir-lite.tex
\begin{table}[!t]
\centering
\setlength\tabcolsep{2.2pt}
\caption{\textbf{Results on BEIR.} The metric is nDCG@10.
The results of the methods marked with $^\dagger$ are from our own re-implementation. 
ST5 GPL denotes Sentence-T5 trained on GPL datasets~\citep{Wang2021GPLGP}.}
\label{table:beir}

\begin{tabular}{@{}l cccccc c @{}}

\toprule
\textbf{Method} & Arg & Covid & NFC & SciFact & SciDocs & FiQA & \textbf{Avg.}\\

\midrule
\multicolumn{6}{@{}l}{\emph{Sparse retrieval}}\\

BM25~\citep{Robertson2009ThePR}
& 29.1  & 58.9 & 33.5  & 67.4 & 14.8 & 23.6 & 37.8
\\

DocT5Query~\citep{Cheriton2019FromDT}
&  34.9  & 71.3 & 32.8 & 67.5 &  16.2 & 29.1 & 41.9\\

\midrule
\multicolumn{6}{@{}l}{\emph{Dense retrieval}}\\
ANCE~\citep{Xiong2020ApproximateNN}
& 31.4  & 73.3 & 23.1 & 50.8 & 12.2 & 29.5 & 36.7
\\

ST5 GPL$^\dagger$~\citep{Ni2021SentenceT5SS} 
& 32.1  & 74.4 & 30.1 & 58.6 & 12.7 & 26.0 & 39.0
\\

GTR-Base~\citep{Ni2021LargeDE}
& 37.3 & 61.2 & 30.0 & 58.4 & 14.0 & 35.1 & 39.3
\\

Contriever~\citep{Izacard2021UnsupervisedDI}
& 40.0  & 68.8 & 33.5 & 61.4 & 16.3 & 30.7 & 41.8
\\

\midrule
\multicolumn{6}{@{}l}{\emph{Generative retrieval}}\\

GENRE$^\dagger$~\citep{Zhuang2022BridgingTG} & 42.5 & 14.7 & 20.0 & 42.3 & \phantom{0}6.8 & 11.6 & 30.0
\\

\textbf{Ours} & 34.3 & 71.8 & 31.6 & 63.9 & 14.9 & 30.2 & 41.1 \\

\bottomrule
\end{tabular}
\end{table}

%% file: tables/ablation-id.tex
\begin{table}[!t]
\centering
\setlength\tabcolsep{3pt}
\caption{Zero-shot evaluation on retrieving new documents with different document tokenization methods.
The second column indicates the type of docid, where BERT-HC denotes BERT-Hierarchical-Clustering~\citep{Tay2022TransformerMA}, Prefix-HC denotes Prefix-aware BERT-Hierarchical-Clustering~\citep{Wang2022ANC}, and dAE denotes discrete auto-encoding.
}
\label{table:aba-id}
\begin{tabular}{@{}l l  c  ccc @{}}
\toprule

& & \textbf{NQ} (R@1) & \multicolumn{3}{c}{\textbf{BEIR} (nDCG@10)}  \\

Method & Docid & Unseen & Arg & NFC & SciDocs \\

\midrule

DSI-Naive$^\dagger$~\citep{Tay2022TransformerMA} & Naive String
& \phantom{0}0.0 & \phantom{0}0.1 & \phantom{0}1.0 & \phantom{0}0.1
\\

DSI-Atomic$^\dagger$~\citep{Tay2022TransformerMA} & Atomic
& \phantom{0}0.0 & \phantom{0}0.2 & \phantom{0}0.8 & \phantom{0}0.1
\\

GENRE$^\dagger$~\citep{DeCao2020AutoregressiveER} & Title
& \phantom{0}6.0 & \phantom{0}0.0 & \phantom{0}2.4 & \phantom{0}0.6
\\

DSI$^\dagger$~\citep{Tay2022TransformerMA} & BERT-HC
& \phantom{0}1.3 & \phantom{0}1.8 & 11.1 & \phantom{0}5.9
\\

NCI~\citep{Wang2022ANC} & Prefix-HC
& 15.5 & \phantom{0}0.9 & \phantom{0}4.3 & \phantom{0}1.2
\\

\midrule

\textbf{Ours} & dAE
& \textbf{34.2} & \textbf{12.1} & \textbf{12.1} & \textbf{12.3}
\\

\bottomrule
\end{tabular}
\end{table}

%% file: tables/speed.tex
\begin{table}[!t]
\centering
\setlength\tabcolsep{5.5pt}
\caption{Efficiency analysis.
}
\label{table:speed}
\begin{tabular}{@{}l cc cc @{}}
\toprule

Method
& Memory & Time (Offline) & Top-$K$ & Time (Online)\\

\midrule

ANCE
& 1160MB & 145min & 100 & 0.69s
\\

GTR-Base
& 1430MB & 140min & 100 & 1.97s
\\

\midrule

\multirow{2}{*}{GENRE}
& \multirow{2}{*}{\textbf{851MB}}  & \multirow{2}{*}{\textbf{0min}} & 100 & 1.41s
\\

&  &  & \phantom{0}10 & 0.69s
\\

\midrule

\multirow{2}{*}{DSI}
& \multirow{2}{*}{851MB} & \multirow{2}{*}{310min} & 100 & 0.32s
\\

&  &  & \phantom{0}10 & 0.21s
\\

\midrule

\multirow{2}{*}{\textbf{Ours}}
& \multirow{2}{*}{860MB} & \multirow{2}{*}{220min} & 100 & \textbf{0.16s}
\\

&  &  & \phantom{0}10 & \textbf{0.10s}
\\

\bottomrule
\vspace{1mm}
\end{tabular}
\end{table}

%% file: tables/case.tex
\begin{table}[!t]
\small 
\centering
\setlength\tabcolsep{4pt}
\caption{Models outputs on NQ320K. The \inyellow{yellow} and \ingray{gray} backgrounds denote the words with higher attention at step $t{=}1$ or $t{=}2$ of \textsc{GenRet}.
\emph{Docid-D} denotes tokenized docid of document D; \emph{Docid-Q} denotes generated docid for query Q.}
\label{table:case}
\begin{tabular}{@{}p{8.5cm}@{}}
\toprule

\textbf{Test query (Q):} what state courts can order a new trial\\

\textbf{Target document (D):} \ingray{United} \ingray{States} \inyellow{appellate} \inyellow{procedure} involves the rules and regulations for \inyellow{filing appeals} in \ingray{state} \ingray{courts} and \ingray{federal} \ingray{courts}. The nature of an appeal can vary greatly depending on the type of case [...]\\
\midrule

-- GENRE: \quad \textbf{Docid-D:} Appellate procedure in the United States; \\ 
\quad \hspace{32pt} \textbf{Docid-Q:} Admission to the Union (\xmark)\\
-- NCI: \quad \hspace{9pt} \textbf{Docid-D:} 22-18-10-1; \hspace{9pt} \textbf{Docid-Q:} 14-10-0-4 (\xmark)\\
-- GenRet: \quad \textbf{Docid-D:} 95-375-59; \hspace{12pt} \textbf{Docid-Q:} 95-375-59 (\cmark)\\

\bottomrule
\vspace*{1mm}
\end{tabular}
\end{table}

%% file: sections/02-RelatedWork.tex
\section{Related work}

\textbf{Sparse retrieval.}
Traditional sparse retrieval calculates the document score using term matching metrics such as TF-IDF~\citep{Robertson1997OnRW}, query likelihood~\citep{Lafferty2001DocumentLM} or BM25~\citep{Robertson2009ThePR}.
It is widely used in practice due to its outstanding trade-off between accuracy and efficiency.
Some methods adaptively assign the term importance using deep neural network~\citep{Zheng2015LearningTR,Guo2016ADR,Dehghani2017NeuralRM}.
With the recent development of pre-trained LMs,
DeepCT~\citep{Dai2019ContextAwareST} and HDCT~\citep{Dai2020ContextAwareDT} calculate term importance using contextualized text representation from BERT.
Doc2Query~\citep{Nogueira2019DocumentEB} and DocT5Query~\citep{Cheriton2019FromDT} predict relevant queries to augment  documents before building the BM25 index using a generative model like T5.
Sparse retrieval often suffers from the lexical mismatches~\citep{Lin2020PretrainedTF}.

\header{Dense retrieval}
Dense retrieval (DR) presents queries and documents in dense vectors and models their similarities with the inner product or cosine similarity~\citep{Karpukhin2020DensePR}.
Compared with sparse retrieval, dense retrieval relieves the lexical mismatch problem.
Various techniques have been proposed to improve DR models, such as hard negative mining~\citep{Xiong2020ApproximateNN,Qu2020RocketQAAO}, late interaction~\citep{Khattab2020ColBERTEA,Santhanam2021ColBERTv2EA}, and knowledge distillation~\citep{Hofsttter2021EfficientlyTA,Lu2022ERNIESearchBC}.
Recent studies have shown the effectiveness of pre-training DR models using contrastive learning on large-scale corpora~\citep{Reimers2019SentenceBERTSE,Ni2021LargeDE,Izacard2021UnsupervisedDI}.
Despite their success, DR approaches have several limitations~\citep{DeCao2020AutoregressiveER,Metzler2021RethinkingSM}:
(i) DR models employ an index-retrieval pipeline with a fixed search procedure (MIPS), making it difficult to optimize the model end-to-end~\citep{Tay2022TransformerMA,Wang2022ANC}.
(ii) Training DR models relies on contrastive learning~\citep{Karpukhin2020DensePR} to 
distinguish positives from negatives, which is inconsistent with large LMs training objectives~\citep{Brown2020LanguageMA} and fails to fully utilize the capabilities of pre-trained LMs~\citep{Bevilacqua2022AutoregressiveSE}.

\header{Generative retrieval}
Generative retrieval is increasing gaining attention. 
It retrieves documents by generating their docid using a generative model like T5.
Generative retrieval presents an end-to-end solution for document retrieval tasks~\citep{Tay2022TransformerMA,Metzler2021RethinkingSM} and allows for better exploitation of the capabilities of large generative LMs~\citep{Bevilacqua2022AutoregressiveSE}. 
\citet{DeCao2020AutoregressiveER} first propose an autoregressive entity retrieval model to retrieve documents by generating titles.
\citet{Tay2022TransformerMA} propose a differentiable search index (DSI) and represent the document as atomic id, naive string, or semantic string.
\citet{Bevilacqua2022AutoregressiveSE} suggest using arbitrary spans of a document as docids.
Additionally, multiple-stage pre-training~\citep{Chen2022CorpusBrainPA,Zhou2022UltronAU}, query generation~\citep{Wang2022ANC,Zhuang2022BridgingTG,Zhou2022UltronAU}, contextualized embedding~\citep{Lee2022ContextualizedGR}, and continual learning~\citep{Mehta2022DSIUT}, have been explored in recent studies.
However, existing generative retrieval models have a limitation in that they rely on fixed document tokenization to produce docids, which often fails to capture the semantic information of a document~\citep{Tay2022TransformerMA}.
It is an open question of how one should define the docids.
To further capture document semantics in docid, we propose document tokenization learning methods.
The semantic docid is automatically generated by the proposed discrete auto-encoding learning scheme in an end-to-end manner.

\header{Discrete representation learning}
Learning discrete representations using neural networks is an important research area in machine learning.
For images, \citet{Rolfe2016DiscreteVA} proposes the discrete variational autoencoder, and VQ-VAE~\citep{Oord2017NeuralDR} learns quantized representations via vector quantization.
Dall-E~\citep{Ramesh2021ZeroShotTG} uses an autoregressive model to generate discrete image representation for text-to-image generation.
Recently, representation learning has attracted considerable attention in NLP tasks, for tasks such as machine translation~\citep{Kaiser2018FastDI}, dialogue generation~\citep{Zhao2018UnsupervisedDS}, and text classification~\citep{Jin2020DiscreteLV,Yu2022LearningST}.
For document retrieval, RepCONC~\citep{Zhan2021LearningDR} uses a discrete representation learning method based on constrained clustering for vector compression.
We propose a document tokenization learning method for generative retrieval, which captures the autoregressive nature of docids by progressive training and enhances the diversity of docids by diverse clustering techniques.

%% file: sections/07-Conclusion.tex
% \vspace*{-2mm}
\section{Conclusions}
This paper has proposed a document tokenization learning method for generative retrieval, named \textsc{GenRet}. 
The proposed method learns to tokenize documents into short discrete representations (i.e., docids) via a discrete auto-encoding approach, which ensures the semantics of the generated docids.
A progressive training method and two diverse clustering techniques have been proposed to enhance the training of the model.
Empirical results on various document retrieval datasets have demonstrated the effectiveness of the proposed method.
Especially, \textsc{GenRet} achieves outperformance on unseen documents and can be well generalized to multiple retrieval tasks.
In future work, we would like to extend the approach to large document collections.
We also plan to explore generative pre-training for document tokenization using large-scale language models.
Additionally, we intend to investigate the dynamic adaptation of docid prefixes for progressive training.

%% file: appendix.tex
% \begin{appendices}

% \appendix

% \end{appendices}